\documentclass[preprint,showpacs,preprintnumbers,amsmath,amssymb]{revtex4}

\usepackage{graphicx}
\usepackage{dcolumn}
\usepackage{bm}
\usepackage{epsfig}

\begin{document}

\preprint{APS/123-QED}

\title{The Parker Magnetostatic Theorem}

\author{B. C. Low\\High Altitude Observatory\\
National Center for Atmospheric Research\\
February 23, 2010}

 
\begin{abstract}

We demonstrate the Parker Magnetostatic Theorem in terms of a small neighborhood in solution space containing continuous force-free magnetic fields in small deviations from the uniform field.  These fields are embedded in a perfectly conducting fluid bounded by a pair of rigid plates where each field is anchored, taking the plates perpendicular to the uniform field.  Those force-free fields obtainable from the uniform field by continuous magnetic footpoint displacements at the plates have field topologies that are shown to be a restricted subset of the field topologies similarly created without imposing the force-free equilibirum condition.  The theorem then follows from the deduction that a continuous nonequilibrum field with a topology not in that subset must find a force-free state containing tangential discontinuities. 

\end{abstract}



\section{Introduction}

We give a new demonstration of the Magnetostatic Theorem of Parker (1994) within the model of Parker (1972) where the theorem was first proposed.  This theorem shows how high electrical conductivity may result in significant resistive dissipation in a hydromagnetic plasma through the spontaneous formation of electric current sheets (Parker 1979, 1994).  As a fundamental explanation of the ubiquitous presence of high-temperature, magnetized plasmas in the astrophysical universe, interest and debates over this theorem have appeared in many publications over the years (Parker 1972, 1979, 1994, 2000, Rosner \& Knobloch 1982, Tsinganos et al. 1984, van Ballegooijen 1985, 1988, Hahm \& Kulsrud 1985, Moffatt 1985, Zweibel \& Li 1987, Ng \& Bhattacharjee 1998, Low 1990, 2006b, 2007, 2010, Bogoyavlenskij 2000, Craig \& Sneyd 2005, Low \& Flyer 2007, Janse \& Low 2009, 2010, Huang et al. 2009, Low \& Janse 2009, Aly \& amari 2010).  The theorem has a clear physical statement but is difficult to express mathematically (Janse, Low \& Parker 2010, hereafter referred to as JLP).  The formidable part of the theory is in the explicit description of the topology of three dimensional (3D) solenoidal fields.  We state the Magnetostatic Theorem in Section 2, analyze in Section 3 some properties involved in boundary displacements of magnetic footpoints, and, finally, demonstrate the theorem in Section 4. We conclude the paper in Section 5.

\section{Statement of the Magnetostatic Theorem}

The induction equation
\begin{equation}
\label{induction}
{\partial {\bf B} \over \partial t} = \nabla \times \left( {\bf v} \times {\bf B} \right)  
\end{equation}
\noindent
describes the evolution of a magnetic field ${\bf B}$ in a perfectly conducting fluid moving with velocity ${\bf v}$.  Magnetic flux is frozen-into the fluid so that the magnetic flux surfaces (MFSs) and field lines (FLs) move as fluid surfaces and lines of fluid particles.  Taking ${\bf v}$ to be continuous in space implies that these geometric surfaces and lines are continuously deformed.  Topology as defined in standard mathematics textbooks deals with the properties of geometric objects that are invariant under completely general, continuous transformations, including transformations that do not preserve any geometric metric.  The fields at two instants of time are the topological transformation of one into the other with certain toplogical properties that do not change.  Ultimately, these properties are physically expressible as the conservation of magnetic flux across fluid surfaces (Low 2006a, LJ).  Given any field ${\bf B}$ there is a set of topological properties we denote formally as $T({\bf B})$ that is invariant in time.  Under the frozen-in condition, $T({\bf B})$ identifies a field physically for all time.

Consider the infinite domain $V : |z| < L_0$, $L_0$ being a constant, in standard Cartesian coordinates. The boundaries $z = \pm L_0$ are perfectly-conducting rigid plates where the field has a fixed normal component,
\begin{equation}
\label{Bz_bc}
B_z |_{z = \pm L_0} = F_{\pm} (x, y) .
\end{equation}
\noindent
The continuity of the tangential electric field at the plate, demanded by Maxwell's equations, imposes the boundary condition
\begin{equation}
\label{v_bc}
{\bf v} |_{z = \pm 1} = 0 ,
\end{equation}
\noindent
assuming that $F_{\pm} \ne 0$.  We limit ourselves to fields that thread across $V$ from one boundary plate to the other, that is, the case of subsystems of flux closed entirely within $V$ is excluded.  Then the unchanging $T({\bf B})$ comprises the mapping of points on one plate to those on the other as pairs of boundary footpoints of FLs, the twist in all definable fluxtubes, and the twists among fluxtubes, topological features easily stated conceptually but difficult to describe in explicit mathematical terms (JLP).  

With a tenuous atmosphere like the solar corona in mind, we adopt the approximation that the non-magnetic forces of the fluid are negligible.  Then the static equilibrium of the magnetized fluid is described by the force-free equations
\begin{eqnarray}
\label{fff_zeroLF}
\left( \nabla \times {\bf B} \right) \times {\bf B} & = & 0 , \\
\label{solenoid}
\nabla \cdot {\bf B} & = & 0 , 
\end{eqnarray}
\noindent
stating that the field exerts no force on the fluid.  Under the frozen-in condition, the field independent of its state has a distinct identity defined by $T({\bf B})$.  So the construction of a force-free field in the above domain is not a classical boundary value problem, but a problem subject to both boundary condition (\ref{Bz_bc}) and a prescribed $T({\bf B})$.   Any expression of this non-classical mathematical problem involves integral equations since topology is a global property (JLP).  The Magnetostatic Theorem states that for most prescriptions of $(F_{\pm} (x, y), T({\bf B}))$, equations (\ref{fff_zeroLF}) and (\ref{solenoid})
have no continuous solutions, which in classical analysis simply means no solutions.  

The problem as posed is nevertheless physically meaningful, and it only means that its solutions must contain magnetic tangential discontinuities, the so-called weak solutions of mathematical analysis (Courant \& Hilbert 1962, JLP).  These are discontinuities of ${\bf B}$ across MFSs subject to continuity of $B^2$ everywhere, so that the integral version of equations (\ref{fff_zeroLF}) and (\ref{solenoid}) is satisfied.  The magnetic tangential discontinuities contain the current sheets and the continuity of $B^2$ ensures that these current-sheet surfaces are in macroscopic equilibrium. 

Let us concentrate on the case of a force-free field described by the infinite series
\begin{eqnarray}
\label{B_series}
{\bf B} (x, y, z) & = & {\bf \hat z} + {\bf b}(x, y, z)  \nonumber \\
                  & = & {\bf \hat z} + \sum_{n=1}^{\infty} \epsilon^n{\bf b}_n (x, y, z) ,
\end{eqnarray}
\noindent
where $\epsilon$ is a constant small parameter, assuming that such a series is analytic with a finite radius of convergence independent of spatial position (Rosner \& Knobloch 1982, JLP).  We may think of some solution space of force-free states in which this series picks out a neighborhood around the uniform field of unit strength.  Some of these states are produced from the uniform field under the frozen-in condition in the manner first considered by Parker (1972).  Starting with the uniform field, first relax the rigid anchoring of the field at $z = \pm L_0$, and then shuffle the magnetic footpoints by some continuous displacement to redistribute the boundary $B_z$ and introduce magnetic twist into the flux tubes in $V$.  Now take the boundary plate as rigid again, freezing the new $B_z$ distribution and the new field topology created.  The Magnetostatic Theorem then states that for most continuous footpoint displacements, the deformed field may not find a continuous force-free equilibrium.  In the context of the infinite series (\ref{B_series}), the claim is that there are footpoint displacements producing deformed fields whose force-free states are not found in the $\epsilon$-neighborhood of the uniform field.  

\section{Field Deformation by Boundary Footpoint Displacements}

We take the footpoint displacements to be continuous. Any discontinuous displacement at the boundary naturally produces a current sheet extending from the boundary into the domain.  Let us use the acronym CMFD for continuous magnetic footpoint displacements.  We treat the properties of CMFDs, and relate them to a property of potential fields studied by Low (2007) and Janse \& Low (2009), hereafter referred to as L07 and JL09.  This prepares us for the demonstration in Section 4.  

\subsection{Field Topology and Footpoint Motions}

Take the $z$ component of the induction equation (\ref{induction}) for a velocity ${\bf v} = {\bf u} = (u_x. u_y, 0)$ on some plane of constant $z$,
\begin{equation}
\label{Bz_induction}
{\partial B_z \over \partial t} + \nabla_{\perp} \cdot \left( B_z {\bf u} \right) = 0 , 
\end{equation}
\noindent
where the subscript $\perp$ indicates partial differentiations in the constant $z$ plane.  If ${\bf u}$ is given, equation (\ref{Bz_induction}) poses an initial value problem for $B_z(x, y, z, t)$ evolving from some initial distribution  $B_z(x, y, z, 0)$ at $t = 0$.  If we know $B_z(x, y, z, t)$, instead, equation (\ref{Bz_induction}) poses a static problem at each instant of time to determine the velocity ${\bf u}$.

Express the  velocity in the general form 
\begin{equation}
\label{u_phi_psi}
{\bf u} = \nabla_{\perp} \phi + \nabla_{\perp} \times \left( \psi {\hat z} \right) , 
\end{equation}
\noindent
separating the compressible irrotational from the incompressible rotational parts, described by $\phi$ and $\psi$, respectively.  If we set $\psi \equiv 0$, the fluid merely compresses the magnetic flux threading across the constant-$z$ plane without twisting the field.  Similarly, if we set $\phi \equiv 0$, the fluid twist the field threading across that plane without compression.  In the latter, $B_z$ can still change with time because of the frozen-in flux transport in the constant-$z$ plane.  To illustrate, set $\nabla_{\perp} \cdot {\bf u} = 0$ to obtain
\begin{equation} 
\label{Bz_incompr}
{\partial B_z \over \partial t} + {\bf u} \cdot \nabla_{\perp} B_z = 0 , 
\end{equation}
\noindent
offering the possibility of twisting the field without changing the $B_z$-distribution by an incompressible rotational displacement with $\psi(x, y, t)$ being a strict function of the unchanging $B_z$.

Consider a given $B_z(x, y, z, t)$, with $B_z(x, y, z, 0) = H_0(x, y)$ at $t = 0$ and $B_z(x, y, z, t_1) = H_1(x, y)$ at $t = t_1$, at two chosen times, keeping in mind we are dealing with events in a constant-$z$ plane.  That evolution can be accounted for by a purely compressive irrotational flow.  Set $\psi \equiv 0$ and then solve 
\begin{equation}
\label{phi}
{\partial B_z \over \partial t} + \nabla_{\perp} \cdot \left( B_z \nabla_{\perp} \phi \right) = 0 , 
\end{equation}
\noindent
as an elliptic partial differential equation for $\phi$.  The solution is unique under suitable physically reasonable boundary conditions for the unbounded constant-$z$ plane.  In other words, the evolution from $B_z = H_0$ at $t = 0$ to $B_z = H_1$ at $t = t_1$ can be achieved through a purely irrotational footpoint displacement.  No magnetic twist is introduced in this process.  Upon arrival at the state $B_z = H_1$ at $t = t_1$, magnetic twist can be put into that field subsequently, if so desired, without changing $B_z$ beyond this point in time.  This is done by an incompressible rotational displacement $\psi = \psi(H_1, t)$ that leaves $B_z$ unchanged in time.

The details of $B_z(x, y, z, t)$ in the period $0 < t < t_1$ determines the precise form of the associated velocity ${\bf u} = \nabla_{\perp} \phi$ in the above process.  There is an infinity of prescribable $B_z(x, y, z, t)$ sharing a pair of initial and final states $(H_0, H_1)$.  Hence, there is an infinity of compressible, irrotational CMFDs that can take $B_z$ from initial to final prescribed distribtions.  Now let us distinguish between the Eulerian velocity of the fluid from the Lagrangian velocity of a fluid particle.  The CMFDs are the Lagrangian displacements of actual fluid particles on the constant-$z$ plane.  Different Eulerian velocities may produce the same final $B_z$-distribution, but their respective actual CMFDs are quite distinct.  In other words, the footpoints on the constant-$z$ plane are displaced continuously in different manners although all the different Eulerian velocities bring about the same final $B_z$-distribution.  It should be emphasized that this is possible only if the footpoints are free to move in the full two-dimensionality of the constant-$z$ plane.  

Returning to the Parker two-plate domain with an initial uniform field, we draw two important conclusions.
Firstly, if the CMFDs at the boundaries $z = \pm L_0$ are incompressible, then it is obvious that the distribution $B_z = 1$ remains at those boundaries.  The field merely acquires a twist.  The different terms in the series (\ref{B_series}) must then satisfy $b_{n,z} = 0$ at $z = \pm L_0$ for all $n$; see the study Low (2010).

If the CMFDs at the boundaries $z = \pm L_0$ are compressible but irrotational, then $B_z$ must change at those boundaries.  Thus, we pose the boundary conditions
\begin{equation}
\label{bnz_bc}
b_{n,z}|_{z = \pm L_0} = f_{n,\pm} (x, y) ,
\end{equation}
\noindent
for $n = 1, 2, 3, ...$, where $f_{n,\pm}$ are prescribed.  In other words, we have the expansion
\begin{equation}
\label{F_f}
F_{\pm} (x, y) = 1 + \sum_{n = 1}^{\infty} \epsilon^n f_{n,\pm} (x, y) ,
\end{equation}
\noindent
to describe the final $B_z$ at the two boundaries produced by an imposed CMFD.  We had simply treated $\epsilon$ as a constant parameter until now.  The free specification of $B_z$ at the boundaries may be taken to be the physical origin of that parameter.

When we displace the footpoints of an initial uniform field at both plates with a rotational motion, i.e., $\psi \ne 0$, it does not mean that we would have twisted the field.  It is the relative motion of the pair of footpoints of a FL that determines whether a twist has been built into the field.  To avoid this qualification, we shall henceforth consider deforming the uniform field with the imposed CMFDs taken only on $z = -L_0$, that is, the magnetic footpoints are not displaced on $z = L_0$ where there is no change in the $B_z$-distribution.  There is no loss of generality to the principal physical point we wish to make. In this case, a rotational CMFD twists the uniform field whereas a compressive, irrotational ($\psi \equiv 0, \phi \ne 0$) CMFD introduces no twist on {\it all} scales.  What is meant by untwisted is in this case unambiguously defined for the field deformed with the latter CMFD.  Moreover, there is an infinity of compressive, irrotational CMFDs that can take the uniform field to deformed states, all sharing the same $B_z$-distributions at the two boundaries but not topologically equivalent by virtue of their distinct footpoint connectivities.

\subsection{The Janse-Low Result}

The possibility of an infinity of untwisted fields in some given domain, all fields sharing the same boundary flux distribution but having different topologies, was encountered in studies L07 and JL, treating the domain between two concentric spheres and the upright cylindrical domain, respectively.  Let us briefly summarize the JL result.

Consider a potential field ${\bf B}_{pot}(L)$ in form of a unidirectional flux entering an upright cylinder of length $L$ and radius $R_0$, from one cylinder end $z = -L$ and exiting at the other end $z = L$.  This potential field is tangential at $R = R_0$, using the usual cylindrical coordinates.   The boundary flux distributions $B_z$, positive definite in this case, at $z = \pm L$ define this potential field uniquely.  By integrating 
\begin{eqnarray}
\label{FL}
{dx \over dz} & = & {B_x(x, y, z) \over B_z(x, y, z)} , \nonumber \\
{dy \over dz} & = &  {B_y(x, y, z) \over B_z(x, y, z)} ,
\end{eqnarray}
\noindent
with ${\bf B} = {\bf B}_{pot}$ we obtain each of its FLs described by $x(z)$ and $y(z)$, two coordinates of a point along the FL in terms of the third $z$ as the independent variable.  These FLs define the map ${\mathcal M} \left[ {\bf B}_{pot}(L) \right] : (x_B, y_B, -L) \rightarrow (x_T, y_T, -L)$:
\begin{eqnarray}
\label{ft_map}
x_T & = & x_B + \int_{z = -L}^{z = L} {B_x\left[x(z), y(z) , z\right] \over B_z\left[x(z), y(z), z\right]} dz , \nonumber \\
y_T & = & y_B + \int_{z = -L}^{z = L} {B_y\left[x(z), y(z) , z\right] \over B_z\left[x(z), y(z), z\right]} dz ,
\end{eqnarray}
\noindent
with ${\bf B} = {\bf B}_{pot}$, relating a footpoint $(x_B, y_B, -L)$ to the footpoint $(x_T, y_T, -L)$.  It was discovered in JL that the map ${\mathcal M} \left[ {\bf B}_{pot}(L) \right]$ is generally dependent on $L$, the exceptions being cases containing special symmetries.  These symmetries include the obvious case of axisymmetry as well as field that are symmetric about the $z = 0$ plane, for examples.  This $L$-dependence of the footpoint map is a 3D effect.

Consider deforming continuously a potential field ${\bf B}_{pot}(L_1)$, along with its domain $L = L_1 \ne L_0$, under the frozen-in condition into a non-potential field ${\bf B}_{df}(L_1;L_0)$ to fit into the domain $L = L_0$, treating the cylinder ends as rigid conductor.  Throughout our discussion we assume no change in the cylinder radius.  In this deformation the footpoint map ${\mathcal M}$ is a topological invariant.  There is no way for the deformed field ${\bf B}_{df}(L_1;L_0)$ to assume the potential state ${\bf B}_{pot}(L_0)$ by some suitable further deformation without also moving the footpoints at the boundary, because their footpoint maps ${\mathcal M}$ are not the same.  With its invariant topology, the deformed field must find some force-free state distinct from ${\bf B}_{pot}(L_0)$ and the arguments in JL concluded that the force-free state must contain tangential discontinuities.

For our deductive analysis here, it is possible to be rigorously specific about a concept of an untwisted field without having to describe its topological properties explicitly.  We call all potential fields, unique in their respective (simply-connected) $L$-domains, untwisted; see JLP for an important discussion of multiply-connected domains.  This is based on the fact that the magnetic circulation vanishes for the potential field along any closed curve in the simply-connected domain, a well known property.  For the field ${\bf B}_{pot} (L) = {\hat z} + {\bf b}$ with $|{\bf b}| << 1$, every flux tube extending from $z = -L_0$ to $z = L_0$ shows no circulation around the approximately straight tube with an axial flux conserved along it.  If this field is deformed and kept within the regime $|{\bf b}| << 1$, magnetic circulations, quantities that are not invariant under the frozen-in condition, can be created in opposite signs along the flux tube.  Conversely,  the opposite circulations found at any one time along a flux tube can be removed by a suitable mutual cancellation of all of them along the tube.  One thing that is forbidden is to deform any of these potential fields to end up with a positive or negative definite along the tube.  

With this definition of untwistedness, a specific $L =L_0$ domain contains an infinity of untwisted fields sharing the same boundary flux distribution, namely, fields that are deformed from the potential fields ${\bf B}_{pot}(L)$ originally occupying domains with $L \ne L_0$ but deformed to fit into the $L = L_0$ domain.  The claim is these deformed fields must find force-free states containing tangential discontinuities.

The deformation of the uniform field in a $L \ne L_0$ domain into a potential field of the form ${\bf B}_{pot} (L) = {\hat z} + {\bf b}$ does not involve rotational footpoint displacements because the magnetic circulation of the potential field vanishes for all closed curves.  The same particular irrotational CMFD can also be applied to the boundaries $z = -L_0$ of the $L_0$ domain containing a uniform field.  The deformed field so produced is topologically identical to ${\bf B}_{pot} (L)$, that is, they have the same footpoint map.  This clarification assures us that the different maps ${\mathcal M}$ of all the different untwisted fields in the $L_0$ domain are produced by irrotational CMFDs.  

\section{The $\epsilon$-Neighborhood of the Uniform Field}

We first solve the magnetostatic equations for the field given by $\epsilon$-series (\ref{B_series}), without consideration of field topology, and then examine the implications of such analytic solutions in terms of the field topologies they exhibit.

\subsection{Perturbational Analysis}

Rewrite the force-free equation (\ref{fff_zeroLF}) in the form
\begin{equation}
\label{fff_tension}
2 \left( {\bf B} \cdot \nabla \right) {\bf B} - \nabla B^2 = 0 ,
\end{equation}
\noindent
and substitute for the field given by series (\ref{B_series}) to first and second orders,
\begin{eqnarray}
\label{1st_eqn}
{\partial \over \partial z} {\bf b}_1 - \nabla b_{1,z} = 0 , \\
\label{2nd_eqn}
2 \left( {\bf b}_1 \cdot \nabla \right) {\bf b}_1  + 2{\partial \over \partial z} {\bf b}_2 - \nabla \left( b_1^2 +  2 b_{2,z} \right) = 0, 
\end{eqnarray}
\noindent
subject to the solenoidal conditions
\begin{equation}
\label{solenoid2}
\nabla \cdot {\bf b}_1 = \nabla \cdot {\bf b}_2 = 0 ,
\end{equation}
\noindent
and boundary conditions (\ref{bnz_bc}) on $b_{1,z}$ and $b_{2,z}$.  No other boundary conditions are needed in this consideration that has not yet introduced the question of field topology.  

The first order equation (\ref{1st_eqn}) subject to the solenoidal condition implies 
\begin{equation}
\label{b1z_Laplace}
\nabla^2 b_{1,z} = 0 ,
\end{equation}
\noindent
which has a unique solution satisfying the prescribed boundary values at $z = \pm L_0$.  Here and elsewhere we assume the far boundary condition that all departures from the uniform field vanish at $x^2 + y^2  \rightarrow \infty$.  The solenoidal condition shows that the first order magnetic field has the form
\begin{equation}
\label{b1}
{\bf b}_1 = \nabla \Phi_1(x, y, z) + {\bf a}_1(x, y) ,
\end{equation}
\noindent
where ${\bf a}_1(x, y) = (a_{1,x}, a_{1,y}, 0)$ is an arbitrary solenoidal vector and $\nabla \Phi_1$ is the unique potential field whose $z$ component accounts for $b_{1,z}$ computed above.  If ${\bf a}_1 \equiv 0$, ${\bf b}_1$ is an untwisted potential field.

First note that the $z$ component of equation (\ref{2nd_eqn}) is an identity whereas the other two components take the forms
\begin{eqnarray}
\label{b2_eqn}
{\partial b_{2,x} \over \partial z} - {\partial b_{2,z} \over \partial x} - \omega \left[
{\partial \Phi_1 \over \partial x} + a_{1,x} \right] & = & 0 , \nonumber \\
{\partial b_{2,y} \over \partial z} - {\partial b_{2,z} \over \partial y} - \omega \left[
{\partial \Phi_1 \over \partial y} + a_{1,y} \right] & = & 0 ,
\end{eqnarray}
\noindent 
where we have expressed ${\bf b}_1$ using equation (\ref{b1}) and introduced $\omega = {\partial a_{1,y} \over \partial x} - {\partial a_{1,y} \over \partial x}$.  Eliminating $b_{2,x}$ and $b_{2,y}$ by using the solenoidal condition in favor of $b_{2,z}$ gives the Poisson equation 
\begin{equation}
\label{b2z_Poisson}
\nabla^2 b_{2,z} + {\partial \over \partial x}\left[ \omega \left(  {\partial \Phi_1 \over \partial x} + a_{1,x} \right) \right] + {\partial \over \partial y}\left[ \omega \left(  {\partial \Phi_1 \over \partial y} + a_{1,y} \right) \right] = 0 .
\end{equation}
\noindent
The solution $b_{2,z}$ subject to its boundary values at $z = \pm L_0$ is unique.  This solution, substituted into equations (\ref{b2_eqn}), determines the $z$-dependent parts of $b_{2,x}$ and $b_{2,y}$ uniquely.  We are free to linearly superpose this second order magnetic field, denoted by ${\bf b}^*_2(x, y, z)$, with an arbitrary solenoidal vector ${\bf a}_2 (x, y) = (a_{2,x}, a_{2,y}, 0)$, and still have a valid solution.  So we express the solution as
\begin{equation}
\label{b2}
{\bf b}_2 = {\bf b}^*_2(x, y, z) + {\bf a}_2 (x, y) .
\end{equation}
\noindent
The free vectors ${\bf a}_1$ and ${\bf a}_2$ describe the twist in the net field (\ref{B_series}) up to second order.  

For example, if ${\bf a}_1 \ne 0$, this solenoidal field by itself has FLs that are the same closed curves on {\it all} constant-$z$ planes.  This property follows from the boundary condition that the perturbations all vanish at $x^2 + y^2 \rightarrow \infty$.  The implied magnetic circulations around the principal flux in the $z$ direction of any flux tube generally cannot mutually cancel. Thus, to first order in $\epsilon$ the FLs wind with a fixed sense about flux tubes directed in the $z$ direction.  An untwisted field corresponds, at first order in $\epsilon$, to ${\bf a}_1 \equiv 0$, in which case ${\bf b}_1 = \nabla \Phi_1$, a pure potential field.   For such a field, $\omega \equiv 0$, and $b_{2,z}$ is potential by equation (\ref{b2z_Poisson}), so that ${\bf b}^*_2(x, y, z)$  is a potential field.  In this case, the free, $z$-independent vector ${\bf a}_2$ imposes a magnetic twist at the second order in $\epsilon$.  Thus, if there is no twist, we must set  ${\bf a}_2 \equiv 0$.  The magnetic field ${\bf B} = 1 + \epsilon{\bf b}_1 + \epsilon^2{\bf b}_2$ is thus potential to second order.  By induction, the perturbations to all orders are all potential and so the net field is potential.

\subsection{Footpoint Displacements}

Substituting a $\epsilon$-series solution into equation (\ref{ft_map}) defines the footpoint map ${\mathcal M}$ of the points on $z = -L_0$ to the points on $z = L_0$.  Expanding in $\epsilon$, we obtain the formal expressions:
\begin{eqnarray}
\label{ftmap_epsilon}
x_T & = & x_B + \sum_{n=1}^{\infty} \epsilon^n \Delta x_n (x_B, y_B)  , \nonumber \\
y_T & = & y_B + \sum_{n=1}^{\infty} \epsilon^n \Delta y_n (x_B, y_B) ,
\end{eqnarray}
\noindent 
where $\Delta x_n$ and $\Delta y_n$ are complicated functions defined through the definite integrals with respect to $z$, straightforward to work out from equation (\ref{ft_map}).  The topological magnetostatic problem of Parker (1972, 1994) is posed by giving the footpoint map ${\mathcal M}$, among other specifications of the topology of the force-free field to be constructed.  In the $\epsilon$-series formulation, this means prescribing the functional forms of $\Delta x_n(x_B, y_B)$ and $\Delta y_n(x_B, y_B)$, and seeking the fields ${\bf b}_n$.  This is obviously a generally intractable problem.  Fortunately, we can derive a clear conclusion by the following particular problems.  

Suppose, we take a potential field in $V$ of the form ${\bf B} = {\hat z} + \epsilon {\bf b}_{pot}$, where ${\bf b}_{pot}$ is a given potential field independent of $\epsilon$ taken to be small so that $|\epsilon {\bf b}_{pot}| << 1$.  What we have is a series solution with ${\bf b}_1 = {\bf b}_{pot}$, and ${\bf b}_n = 0$, for $n \ge 2$.  In other words, the series (\ref{B_series}) has only a finite number of terms, just 2 terms.  Equations (\ref{ftmap_epsilon}) giving the footpoint map ${\mathcal M}$ remain a pair of infinite series.  Given ${\bf b}_{pot}$, each terms of these two infinite series can be computed explicitly.  Turning the problem around, specifying this footpoint map ${\mathcal M}$ to pose the topological problem leads to its solution ${\bf B} = {\hat z} + \epsilon {\bf b}_{pot}$.  This is a case of a deformed field whose prescribed field topology allows it to attain a force-free state, namely, ${\bf b}_{pot}$.

We can also carry out this computation for a footpoint map ${\mathcal M}$ belonging to a deformed field ${\bf B} = {\hat z} + \epsilon {\bf b}_{df}$, not in equilibrium, obtained as followings.  Construct first a potential field ${\bf b}_{pot,L}$ in a domain $V(L): |z| < L$ with $L \ne L_0$, subject to $B_z$-distributions on $z = \pm L$ that are identical to the $B_z$-distributions of ${\bf b}_{pot}$ on 
$z = \pm L_0$.  Compute the footpoint-map functions $\Delta x_n(x_B, y_B)$ and $\Delta y_n(x_B, y_B)$.  Now continuously deform ${\hat z} + {\bf b}_{pot,L}$ along with the domain $V(L): |z| < L$, to obtain ${\bf b}_{df}$ that fits into $V(L_0)$.  The footpoint-map functions $\Delta x_n(x_B, y_B)$ and $\Delta y_n(x_B, y_B)$ remain unchanged and the field remains untwisted.  In demanding that this deformed field be further deformed under the frozen-in condition to be force free, we conclude from our perturbational analysis that it cannot find a continuous solution in the $\epsilon$-neighborhood of the uniform field.  If it could, that force-free state has to be the unique potential field ${\bf b}_{pot}$. That is not possible because the two fields have different footpoint maps.  This demonstrates the Parker Magnetostatic Theorem.

\section{Conclusion}

We have returned to the model of Parker (1972) to demonstrate the Magnetostatic Theorem described in generality in Parker (1994).  The general theory is concerned with the infinite set of field topologies ${\bf \mathcal T}_{fff}$ of all continuous force-free fields admissible in a physical system.  This set is generally a true subset of the set ${\bf \mathcal T}_{all}$ of the topologies of non-force-free continuous fields admissible in the same system, that is, ${\bf \mathcal T}_{fff} \subset {\bf \mathcal T}_{all}$.  Thus a non-force-free, continuous field with a topology not found in ${\bf \mathcal T}_{fff}$ must develop tangential discontinuities as the only way to become force-free as it perserves its topology under the frozen-in condition.  

We have demonstrated the Magnetostatic Theorem by treating the untwisted fields admissible in a physical system.  The problem is simpler than the one posed by twisted fields.  In 3D systems, the set ${\bf \mathcal T}_{all}^{untwd}$ of topologies of untwisted non-force-free fields is generally infinite.  In contrast, a 2D field with no neutral points; for example, an axisymmetric field, has a unique footpoint map, see JL and JLP.  There is only one force-free state for an untwisted field in the 3D system, the unique potential field.  Hence the set of untwisted topologies ${\bf \mathcal T}_{fff}^{untwd}$ realizible in a continuous force-free state is a set with a single member, and, obviously, ${\bf \mathcal T}_{fff}^{untwd} \subset {\bf \mathcal T}_{all}^{untwd}$, thus demonstrating the theorem.  

A similar demonstration was made in the studies L07 and JL, but those studies depend on arguing around topological ideas that are extremely difficult to pin down mathematically.  In the present paper, the study of the force-free fields in the $\epsilon$-neighborhood of the uniform field, treated by Parker (1972), simplifies the general problem.  It is the ordering of the terms by magnitude in the series expansion, beginning with the dominant uniform field, that allows a clear sorting out between what is an untwisted field from those that are twisted.  Basic to this analysis is the property that different footpoint connectivities can be created by footpoint displacements imposed on a uniform field, without twisting the field but all leading to the same boundary flux distribution.  

Counter arguments about the result in JL, like those recently presented by Aly \& Amari (2010), need to be reexamined in light of this new result.  We mention the following open questions that qualify the new result.  The fact that we could not find continuous solution in the $\epsilon$ neighborhood of the uniform field does not preclude the existence of the desired continuous solutions represented by some other small-parameter expansions; see the interesting development of van Ballegooijen (1985) and the discussions in Low (1990, 2010), Parker (1994), and JLP.  Another possibility is that, despite the small-parameter nature of the CMFDs expressed by equation (\ref{ftmap_epsilon}), continuous force-free solutions may exist in $V$ that do match the prescribed footpoint displacements but can only be attained via deformations of finite amplitudes inside the domain.  These questions should not detract from the central feature of our result, that the continuous force-free fields in the $\epsilon$-neighborhood of the uniform field are topologically of a more restricted variety than the non-force-free fields.  The Parker Magnetostatic Theorem makes the fundamental point that this hydromagnetic feature is general.  
 
The National Center for Atmospheric Research is sponsored by the National Science Foundation.

\end{document}